\documentclass[12pt]{article}
\usepackage{epsfig}

\def\beq{\begin{equation}}
\def\eeq#1{\label{#1}\end{equation}}
\def\eeqn{\end{equation}}

\def\beqa{\begin{eqnarray}}
\def\eeqa#1{\label{#1}\end{eqnarray}}
\def\eeqan{\end{eqnarray}}

\let\bar=\overbar

\def\Dslash{\not{\hbox{\kern-4pt $D$}}}
\def\dslash{\not{\hbox{\kern-2pt $\del$}}}

\def\msb{{\bar{\ssstyle M \kern -1pt S}}}

\def\Title#1{\begin{center} {\Large {\bf #1} } \end{center}}

\begin{document}

\Title{Measurements of the CKM angle $\gamma$ using $B^+ \to D K^+$ decays at LHCb}

\bigskip\bigskip


\begin{raggedright}  

{\it Sneha Malde, on behalf of the LHCb collaboration,\index{Reggiano, D.}\\
Department of Physics\\
University of Oxford\\
OX1 3RH,  UK\\
}
Proceedings of the CKM 2012, the 7th International Workshop on the CKM Unitarity Triangle, University of Cincinnati, USA, 28th September - 2 October 2012 

\bigskip\bigskip
\end{raggedright}

\begin{abstract}

Of the three angles that make up the CKM matrix, the least well known is $\gamma$. A precision measurement of this quantity is highly desirable as it forms one of the arbitrary parameters in the Standard Model. Moreover, this is the one angle of the CKM triangle that can be determined in channels that occur via tree-level decays. While loop-level processes could have sensitivity to physics beyond the standard model, tree level processes are expected to be unaltered. Hence a measurement of $\gamma$ in tree level processes leads to a standard model benchmark measurement against which other loop-driven measurements can be compared. The measurements described here are the latest developments from LHCb involving the decay $B^+ \to D K^+$, where the $D$ meson is either a $D^0$ or a $\overline{D^0}$ and the final state of the $D$ meson is accessible from either flavour state.
\end{abstract}

\section{Introduction and methodology}
 The sensitivity to $\gamma$ in $B \to D K$ modes arises from the interference of $b \to c $ and $b \to u $ transitions. Decays studied are classified by the final state of the $D$ meson with naming conventions defined from the original suggestion of their use. They are
\begin{itemize}
\item GLW~\cite{GLW} -- Decays of the $D$ to a CP eigenstate: $D \to KK$, $D \to \pi\pi$
\item ADS~\cite{ADS} -- Decays that are accessible either by a Cabibbo--favoured decay or the doubly-Cabbibo-suppressed decay: $D \to K\pi$, $D \to K\pi\pi\pi$.
\item GGSZ~\cite{GGSZ} -- Decays of self conjugate modes: $D \to K_S^0 \pi\pi$, $D \to K_SKK$.
\end{itemize}
In the case of the GLW and ADS modes the observables with sensitivity to $\gamma$ can be constructed by taking the ratio of supressed and favoured yields or a measurement of the asymmetry between $B^+$ and $B^-$ to a particular final state. These observables are dependent on three common physics parameters which are $\gamma$, $\delta_B$--the strong phase of the $B$ decay, and $r_B$--the ratio of the amplitudes for the decays $B \to D^0 K$ and $B \to \overline{D^0} K$. In the case of the ADS modes there is also dependence on the $D$ decay parameters. As the multibody $D \to K\pi\pi\pi$ decay is treated inclusively it is also necessary to modify the standard ADS dependencies with the inclusion of the coherence factor~\cite{coherencepaper}. Further information on the observables and their sensitivity to $\gamma$ is given in ~\cite{GLW,ADS,twobody,fourbody,combo}.

\begin{figure}[htb]
\begin{center}
\epsfig{file=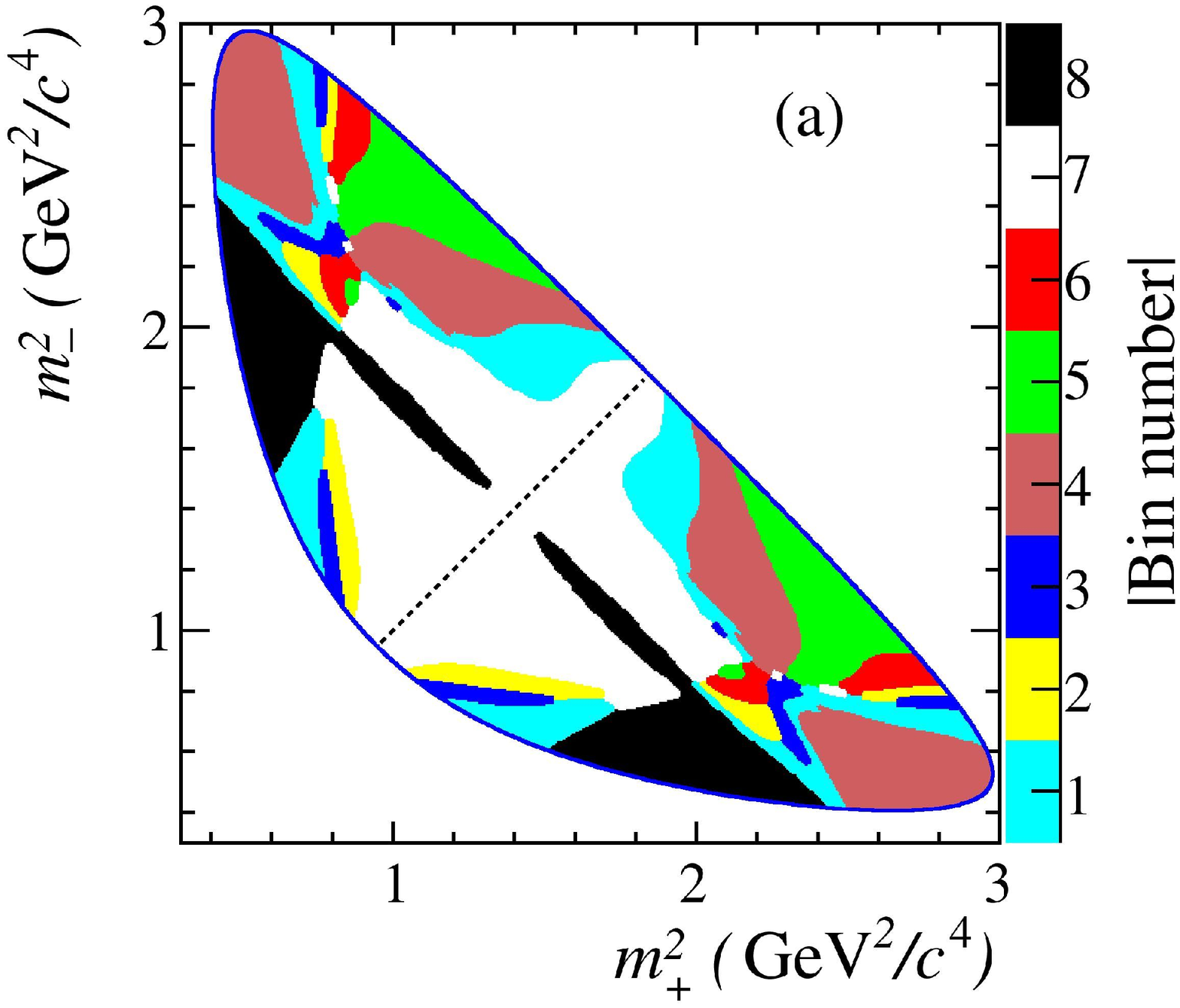,height=2.1in}
\epsfig{file=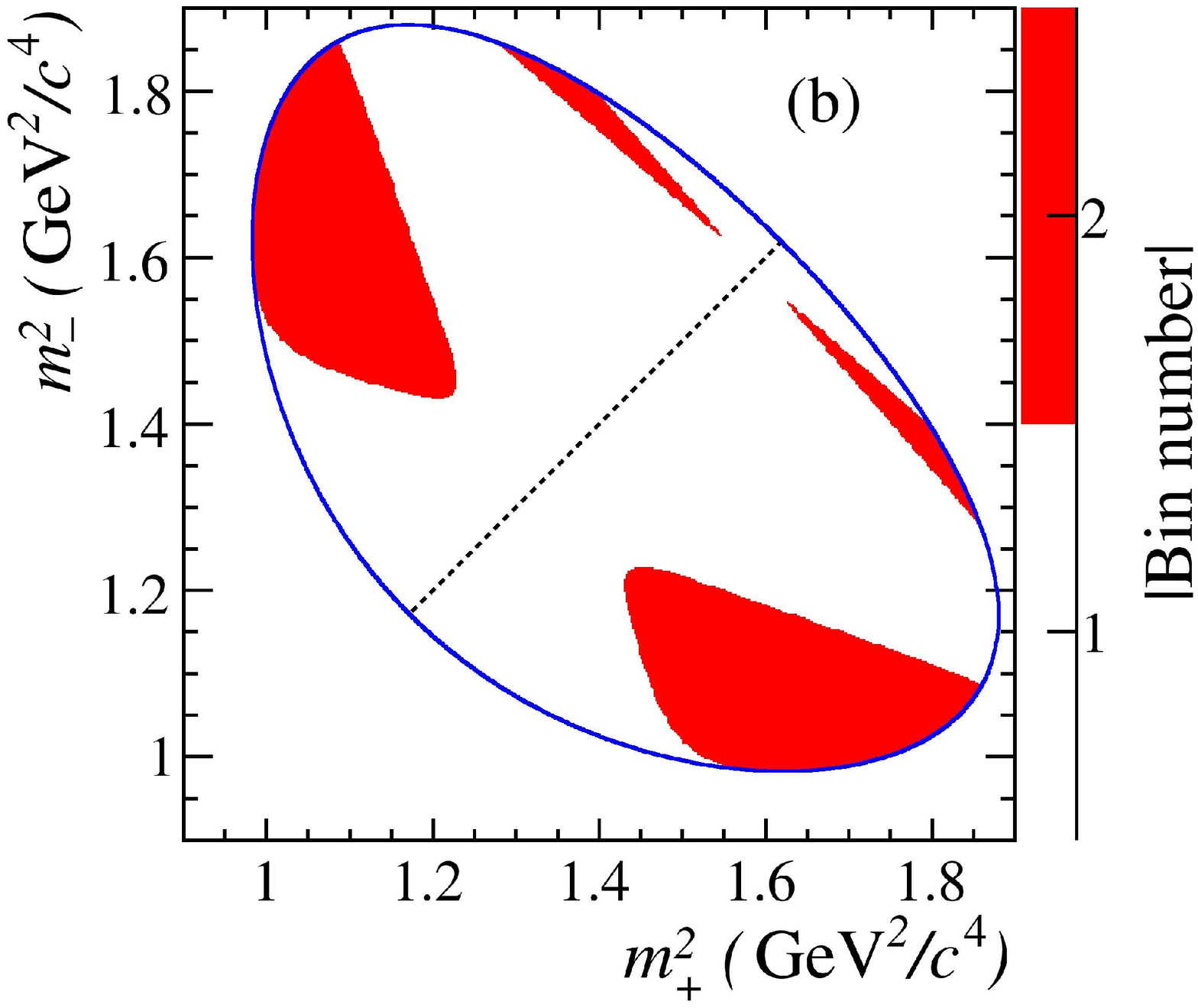,height=2.1in}
\caption{The Dalitz plot binning schemes for $D^0 \to K_S^0\pi\pi$ (left) and $D^0 \to K_S^0 KK$ (right)}
\label{fig:bins}
\end{center}
\end{figure}

In the case of the GGSZ modes, an inclusive treatment leads to almost no $\gamma$ sensitivity. Instead the $D$ decay is analysed on the Dalitz plot which takes into account the varying strong phase of the $D$ decay. A model--independent approach is taken where the Dalitz plot is split into bins. The yield observed in each bin is dependent on the parameters $x_{\pm}$= $r_B \cos(\delta_B \pm \gamma)$, $y_{\pm}= r_B \sin(\delta_B \pm \gamma)$ and $c_i$ and $s_i$, where $c_i$ and $s_i$ are the amplitude--weighted average cosine and sine of the strong phase difference between the $D^0$ and $\overline{D^0}$ decay over a particular bin. The $c_i$ and $s_i$ have been measured directly using charm threshold data at CLEO~\cite{cisi}. The binning schemes used are shown in Fig.~\ref{fig:bins}.

The analyses consider 1.0 fb$^{-1}$ of data collected at LHCb~\cite{detector} during the 2011 run. The exact selection differs depending on the different $D$ final states. Selection criteria are focused on those which reduce the combinatorial background. The variables chosen to do this are topological in nature and depend on quantities such as track impact parameters or particle flight distances, or the fit quality of the secondary and tertiary vertices. Additional selection criteria remove peaking backgrounds which differ depending on the $D$ decay. The Cabibbo--favoured $B \to D \pi$ modes are also analysed and used in determining the mass model via simultaneous fitting where the $B \to D K$ mass distribution shares characteristics with the higher statistics $B \to D \pi$ decay. In the GGSZ modes the $B \to D \pi$ channel is also used to determine the variation in reconstruction efficiency over the Dalitz plot.

\section{ADS and GLW modes}

\begin{figure}[htb]
\begin{center}
\epsfig{file=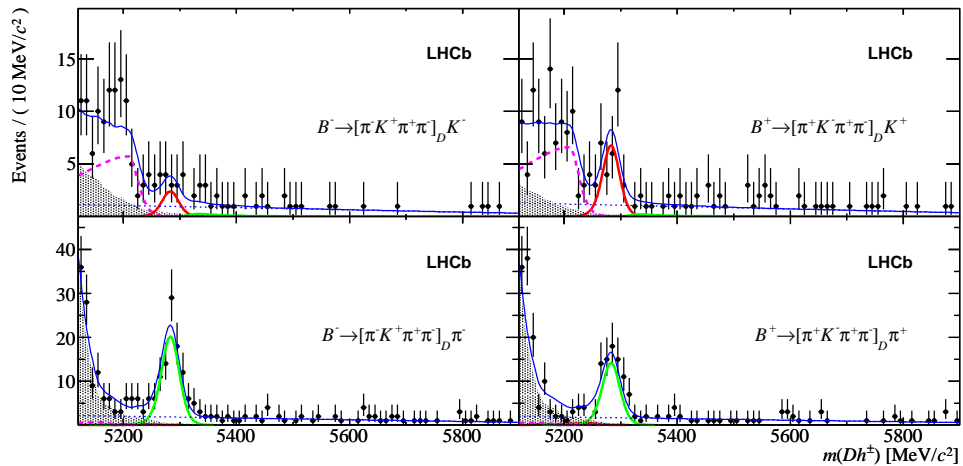,height=2.4in}
\caption{Mass distribution of the suppressed -ADS $D \to K\pi\pi\pi$ decay for $B \to D K$ and $B \to D \pi$ shown separately for $B^+$ and $B^-$}
\label{fig:k3pi}
\end{center}
\end{figure}

The most recent results relate to the $D \to K\pi\pi\pi$ channel. The suppressed decay (denoted by the kaon from the $B$ and $D$ decay having opposite sign) has not previously been observed. The resulting mass distributions for this decay after selection are shown in Fig~\ref{fig:k3pi}. After summing together contributions from $B^+$ and $B^-$ the significance of the suppressed decay in the $B \to D K$ ($B \to D \pi$) channel exceeds 5$\sigma$ (10$\sigma$). This represents the first observation of suppressed-ADS decay in this decay. The asymmetry in $B \to D K$ is calculated to be 0.42$\pm$0.22 where the uncertainty is dominated by the statistical uncertainty. The ratio between the suppressed and favoured decay is 0.0124$\pm$0.0027. Further results are given in Ref.~\cite{fourbody}. The results of the two body decay modes give the first observation of CP violation in $B \to D K$ decays, the first observation of the suppressed-ADS $B \to DK$, $D\to K\pi$ decay with a significance greater than 10$\sigma$, and the measurement of the asymmetries and ratios in the GLW and ADS decay modes~\cite{twobody}. These measurements will all contribute to a combined measurement of $\gamma$.

\section{GGSZ modes and constraints on $\gamma$}

\begin{figure}[htb]
\begin{center}
\epsfig{file=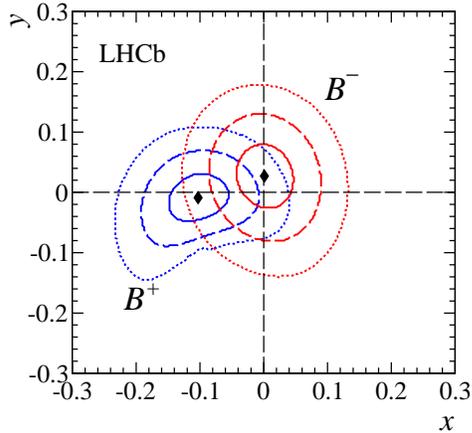,height=2.4in}
\caption{One (solid), two (dashed) and three (dotted) standard deviation confidence levels for  $(x_+,y_+)$ (blue) and $(x_{-}, y_{-})$ (red) as measured in $B^\pm \to D K^\pm$ decays (statistical only). The points represent the best fit central values.}
\label{fig:lik}
\end{center}
\end{figure}

In the GGSZ modes, 690 $B \to D K$, $D \to K_S^0 \pi\pi$ and 110 $B \to D K$, $D \to K_S^0KK$ signal events are observed. The data is then split into bins on the Dalitz plot and the mass distribution in each bin is simultaneously fit to extract the best fit values for $x_{\pm}$ and $y_{\pm}$. The best fit values are~\cite{kspipipap}
\vspace*{-0.4cm}
\begin{equation}
x_-   =    (0.0 \pm 4.3 \pm 1.5 \pm 0.6) \times 10^{-2}, \; \, y_-  = (2.7 \pm 5.2 \pm 0.8 \pm 2.3) \times 10^{-2}, \nonumber\\
\end{equation}
\vspace*{-0.6cm}
\begin{equation}
x_+   =    (-10.3 \pm 4.5 \pm 1.8 \pm 1.4) \times 10^{-2}, \; \,  y_+  = (-0.9 \pm 3.7 \pm 0.8 \pm 3.0) \times 10^{-2}, \nonumber
\end{equation}
 where the first uncertainty is statistical, the second is the systematic uncertainty related to fitting methods and detector effects and the last uncertainty is that related to the precision on the $c_i$ and $s_i$ inputs from CLEO. This is the first time that the $D \to K_S^0KK$ decay mode has been analysed using the model--independent method. The likelihood scan showing the 1, 2, and 3 $\sigma$ contours for the statistical uncertainty is shown in Fig~\ref{fig:lik}. The expected signature for a sample that exhibits $CP$-violation is that the two vectors defined by the coordinates $(x_-,y_-)$ and $(x_+,y_+)$ should both be non-zero in magnitude, and have different phases.  The data show this behaviour, but are also compatible with the no $CP$-violation hypothesis. The results for $x_\pm$ and $y_\pm$ can be interpreted in terms of the underlying physics parameters $\gamma$, $r_B$ and $\delta_B$.  This is done with a frequentist approach with Feldman-Cousins ordering, using the same procedure as described in Ref.~\cite{Belle}. The solution for the physics parameters has a two-fold ambiguity, $(\gamma, \delta_B)$ and $(\gamma + 180^\circ, \delta_B +180^\circ)$.  Choosing the solution that satisfies $0 < \gamma < 180^\circ$  yields $r_B = 0.07 \pm 0.04$, $\gamma = (44^{\,+43}_{\,-38})^\circ$ and $\delta_B = (137^{\,+35}_{\,-46})^\circ$.  The value for $r_B$ is consistent with, but lower than, the world average of results from previous experiments~\cite{PDG}. This low value means that it is not possible to use the results of this analysis, in isolation, to set strong constraints on the values of $\gamma$ and $\delta_B$, as can be seen by the large uncertainties on these parameters. However this analysis remains important as it helps resolves ambiguities when combined with ADS and GLW style results.

\section{Summary}

Using 1.0 fb$^{-1}$ of data LHCb has observed $CP$-violation in $B \to D K$ decays using the two body decays of the $D$ meson. The first observation of the suppressed ADS modes in $D \to K\pi$ and $ D \to K\pi\pi\pi$ decays and the measurement of observable related to $\gamma$ have been made. A model--independent analysis of the GGSZ modes $D \to K_S^0 \pi\pi$ and $D \to K_S^0KK$ has been performed and can be used to set some constraints on the CKM angle $\gamma$. The power of all these measurements is realised in the combination of them all to determine the best fit value for $\gamma$~\cite{combo}.



\end{document}